\documentclass{article}
\usepackage{graphicx} 
\usepackage{ragged2e}
\usepackage{subfig}
\usepackage{geometry}
\usepackage{authblk}

\title{A comparative data study on dinosaur, bird and human bone attributes - A supporting study for convergent evolution.}
\author[1]{Akshita Patil}
\affil[1]{Puna International School, Ahmedabad, India}
\author[2]{Nishchal Dwivedi \thanks{nishchal.dwivedi@nmims.edu}}
\affil[2]{Department of Basic Science and Humanities, SVKM’s NMIMS Mukesh Patel School of Technology Management \& Engineering, Mumbai, India }

\begin{document}

\centering 
\maketitle

{\large Abstract}

\begin{justify}

For over 165 million years, dinosaurs reigned on this planet. Their entire existence saw variations in their body size and mass . Understanding the relationship between various attributes such as femur length, breadth; humerus length, breadth; tibia length, breadth and body mass of dinosaurs contributes to our understanding of the Jurassic era and further provides reasoning for bone and body size evolution of modern day descendants of those from the Dinosauria clade. The following work consists of statistical evidence derived from an encyclopedic data set consisting of a wide variety of measurements pertaining to discovered fossils of a particular taxa of dinosaur. Our study establishes linearly regressive correspondence between femur and humerus length and radii. Furthermore, there is also a comparison with terrestrial bird bone lengths, to verify the claim of birds being closest alive species to dinosaurs. An analysis into bone ratios of early humans shows that terrestrial birds are closer to humans than that of dinosaurs. Not only on one hand it challenges the closeness of birds with dinosaurs, but on the other hand it makes a case of convergent evolution between birds and humans, due to their closeness in regressive fits.

A correlation between bone ratios of dinosaurs and early humans also advances understanding in the structural and physical distinctions between the two species. Overall, the work contains evaluation of dinosaur skeletons and promotes further exploration and research in the paleontological field to strengthen the conclusions drawn thus far.
\end{justify}

\begin{flushleft}
\section{Introduction}
\end{flushleft}
\begin{justify}
Long forgotten members of the animalia kingdom belonging to the dinosauromorpha clade, dinosaurs; still carry around a lot of unanswered questions\cite{benson2014rates,inbook,diogo2017dinosaurs} that one can only theorise about. There are sub-clades within this group of reptiles as well : Dinosauria and dinosauromorpha. The former refers to dinosaurs which have significant evolutionary links to crocodilians and the latter having noticeable evolutionary links to birds \cite{ezcurra2021early}. Dinosaurs were present on the Earth for 165 million years and came into existence about 245 million years ago (mya). They lived during the Mesozoic era, first seen in the middle of the Triassic age and last seen in the late Cretaceous. The dinosauria clan is divided into 2 major groups - Ornithischia and Saurischia. Saurischia further consists of 2 major subgroups  - Sauropodomorpha and Theropoda\cite{currie1997encyclopedia} \cite{langer2010origin}. 

All known data, information and conclusions about dinosaurs is made from their fossils. Fossils are skeletal remains of species that have accidentally been preserved by being buried within silt, sand and sediments. Fossils are studied by expertly extracting the fossil from the matrix it is embedded in, stabilising it using adhesives and consolidants and then creating casts and moulds to identify origin, estimate size, shape and establish a timeline. Scientists study extremely fragmented and fragile bones and hence data is not considered 100\%
 accurate and is used more as estimations. 

The aim of the following study is to answer two questions:

        i) Whether there is a symmetry or fixed ratio between the length, thickness, and radius of various dinosaur bone types. 

        ii) If there is an undiscovered relation between human bone ratios and dinosaur bone ratios. Further extending this question to dinosaur descendants, that is, birds; who are referred to as the last dinosaurs \cite{chiappe2009downsized}. 

This work verifies these results by carrying out computational statistics and looking specifically for linear relationships between lengths and radii of the femur, tibia and humerus bones. We represnet these results using graphs between the estimated body mass (in tonnes)  and femur, humerus bone lengths. Furthermore, the same procedures are carried out for human bone measurements and be used to draw conclusions (if any) between dinosaurs and modern-day humans. These attributes and comparison are then be extended to birds. 

The analysis of specifically limb bones is done due to them being attributed as a promising field of investigation as the genetics of limb formation and growth can be credited for variations in shape proportion and a number of other physiological elements.\cite{diogo2017dinosaurs}
\end{justify}
\begin{flushleft}
\section{Discussion and Results}
\subsection{Comparison across 3 dinosaur subgroups based on skeletal parameters}
\end{flushleft}
\begin{justify}
A closer look into the dinosaur subspecies: 
Sauropodomorphs\cite{currie1997encyclopedia} first evolved in the early Jurassic epoch and persisted on the Earth till the late Cretaceous period, amassing almost 130 million years as a species on this planet. They are typically identified by their leaf-shaped tooth crowns, a small head, and a neck that is at least as long as the trunk of the body and longer than the limbs. Some common examples of this species are the brachiosaurus and the diplodocus. The Sauropodomorpha are descendants of their early ancestor Prosauropods.  Later evolutions were referred to as Sauropods. The Prosauropods are generally stocky in appearance, with hind legs which were longer than their forelimbs. They were five-toed with a long tail which helped the dinosaur balance while walking and running. \cite{carrano2005evolution}

Theropods\cite{currie1997encyclopedia} are part of dinosaur subclade Theropoda and a sister species to the Sauropods (under the Saurischian order). They belong to the Saurischian dinosaur order alongside Sauropodomorphs. A distinct feature of Theropods is that they are primarily bipeds, that is, they rely on their hind limbs for motion. They are slender,long-legged and are theorized to move faster than herbivore dinosaurs. Their lineage is almost entirely known for carnivorous dinosaurs\cite{currie1997encyclopedia}. 

Ornithischians\cite{currie1997encyclopedia} were mainly herbivorous dinosaurs that inhabited the land roughly 200 million years ago until the ultimate extinction of all dinosaur species at the end of the Cretaceous period. Ornithischia, otherwise translated to bird-hipped, consists of dinosaurs such as the triceratops and stegosaurus. Referred to as bird-hipped due to their hip bones being arranged similar to those of birds, that is, it was down-sloping and pointed towards the tail instead of pointing forward as seen in other reptilians/dinosaurs. This was also known as a caudally oriented pubis.\cite{shackelford2017encyclopedia}
Common genetic trends seen in all under the Ornithischian order include an extra bone at the tip of the jaw and foliated teeth(broad and flat teeth). Other variations seen in different lineages within this order include a network of bony tendons that intersected the protrusions of their vertebral column (as seen in \textit{Stegosauruses}) and evolved jaws that were stronger and consisted of several rows of teeth (for example, the \textit{Hadrosaur}). 

The bones considered are the femur which form a portion of the hind leg. In dinosaurs the femur is crowned by a femoral head which forms a right angle with the (femoral) shaft\cite{brusatte2012dinosaur}. Followed by comparison in terms of the humerus, the largest bone in the forelimb \cite{brusatte2012dinosaur}, and the tibia, present in the lower leg adjacent to the fibula.
\end{justify}
\begin{flushleft}
    \subsubsection{Femur circumference with respect to length}
\end{flushleft}

\begin{figure}

\begin{minipage}{.33\linewidth}
\centering
\subfloat[]{\label{main:a}\includegraphics[scale=.3]{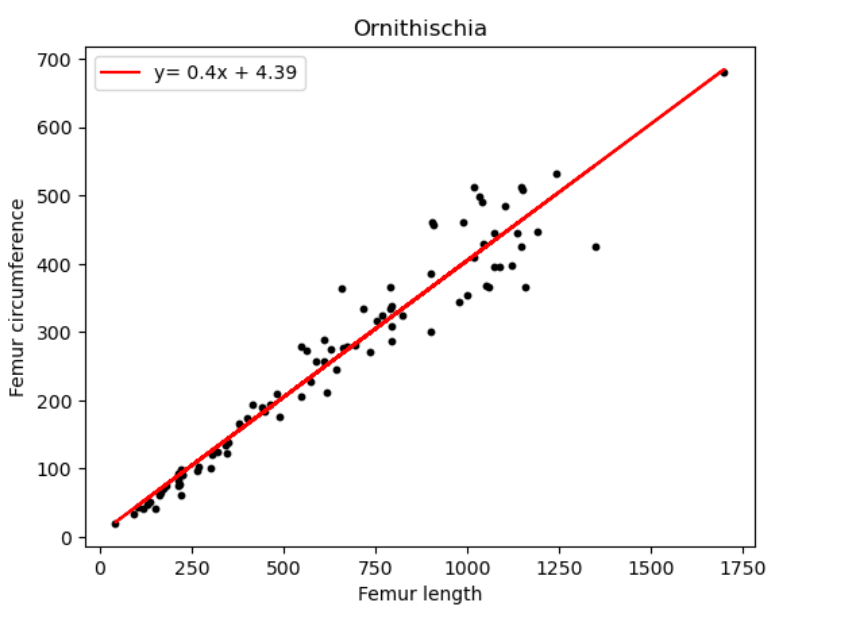}}
\end{minipage}%
\begin{minipage}{.33\linewidth}
\centering
\subfloat[]{\label{main:b}\includegraphics[scale=.3]{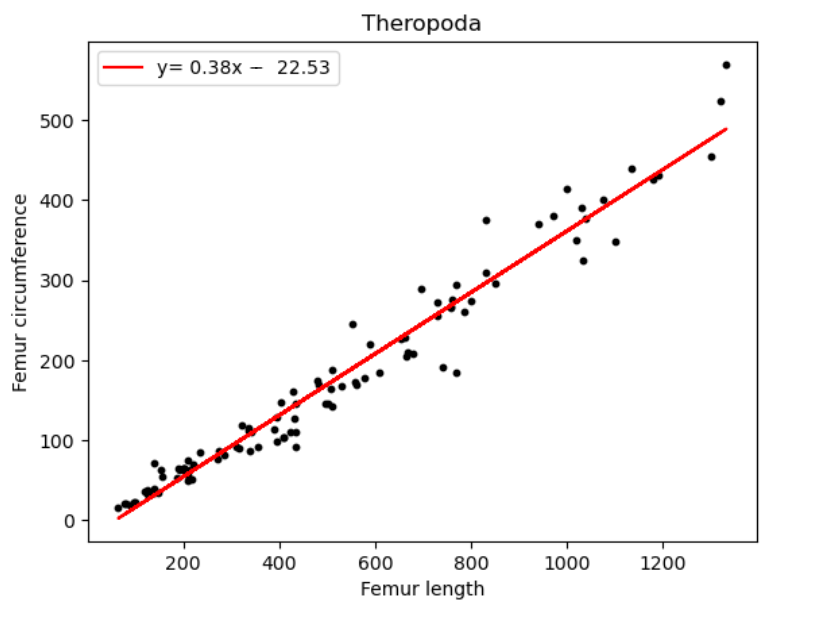}}
\end{minipage}
\begin{minipage}{.33\linewidth}
\centering
\subfloat[]{\label{main:c}\includegraphics[scale=.3]{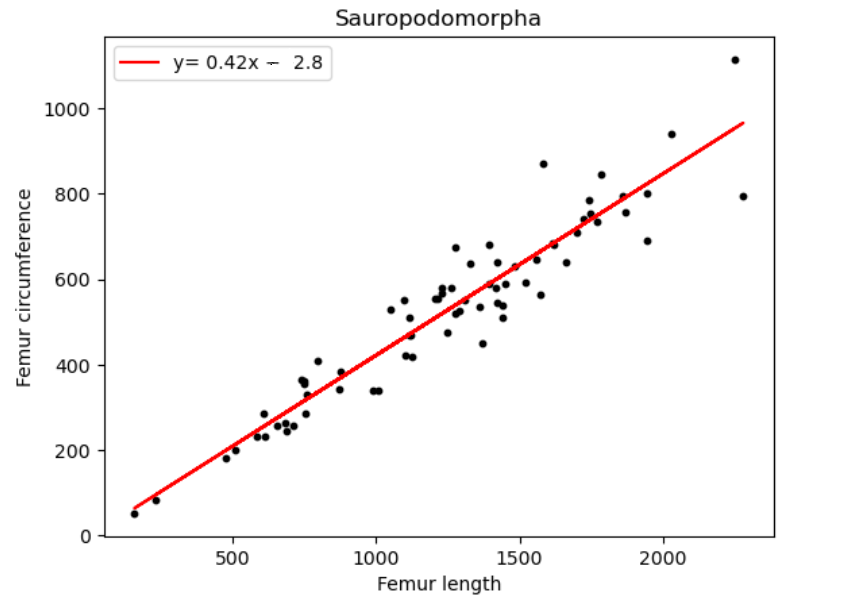}}
\end{minipage}
\caption{(a) Femur circumference vs. length for Ornithischia (b) Femur circumference vs. length for Theropoda (c) Femur circumference vs. length for Sauropodomorpha}
\label{fig:main1}
\end{figure}

\begin{table}[ht]
    \centering
    \begin{tabular}{|c|c|c|}
        \hline
        Name of species/group & Slope of line (regression coefficient) & Regression intercept \\
        \hline
         Sauropodomorpha & 0.42 & -2.8 \\
        
       Theropoda & 0.38 & -22.53 \\
       
        Ornithischia & 0.40 & 4.39 \\
        \hline
    \end{tabular}
    \caption{Slope and coefficient values for femur circumference vs. length}
    \label{tab:simple1}
\end{table}

\begin{justify}
Reviewing figure \ref{fig:main1} and table \ref{tab:simple1} for the dinosaur groups under the Saurischian order. We observe a significant variation in the regression coefficients which implies no comparable ratio was followed between the 2 groups. On considering all 3 groups of dinosaurs we notice that they are all around the same with only a 0.2 difference between each. 
We are also able to conclude that over their period of existence the Sauropodomorphs saw a 0.42 increase in femur circumference with respect to femur length. Followed by Ornithischians who saw a 0.4 increase and lastly Theropods with a 0.38 increment. 

For Sauropods, we can also associate this change in bone length with an increase in body size. This is consistent with previous findings about the body size evolution of the Sauropods starting in the early Jurassic which eventually led to them being some of the largest animals to be recorded to date. \cite{sander2011biology} The earliest dinosaur present in the data is the \textit{Saturnalia Tupiniquim}, dating back to the early Carnian age (nearly 230 mya), it was a small dinosaur estimated to weigh only about 10kg and 1.5 meters in length \cite{langer1999sauropodomorph}. The latest species found were the \textit{Jainosaurus septentrionalis}, \textit{Magyarosaurus dacus} and \textit{Rapetosaurus krausei} of which the \textit{R. krausei} has an estimated length of 15 meters and weighed about 40 kg.\cite{montague2006estimates} It was also noted that the \textit{S. tupiniquim} was bipedal as opposed to subsquent generation Sauropods such as the \textit{R. krausei} being quadrupedal. We believe this can be attributed to an increase in body mass, which led to a slow evolutionary shift from only hind leg locomotion to use of all 4 limbs. 

As for Theropods they seem to show the least change in femur circumference with the passage of time. This can be connected to a speculated evolution in Theropods which is said to have taken in place in the Mesozoic age. Although the evolution does not account for an increase in femur length rather change in the tibia, fibula and metatarsal length is more prominently recorded.\cite{currie1997encyclopedia}

In Ornithischians there is a 0.4 increase in femur circumference in correlation to length. However, no strong links can be established for increase in femur length with pedal transition or body size evolution.\cite{barrett2017evolution}\cite{persons2020anatomical}
\end{justify}
\begin{flushleft}
        \subsubsection{Humerus circumference with respect to length}
    \end{flushleft}
    \begin{justify}
    \begin{figure}

\begin{minipage}{.33\linewidth}
\centering
\subfloat[]{\label{main:a}\includegraphics[scale=.3]{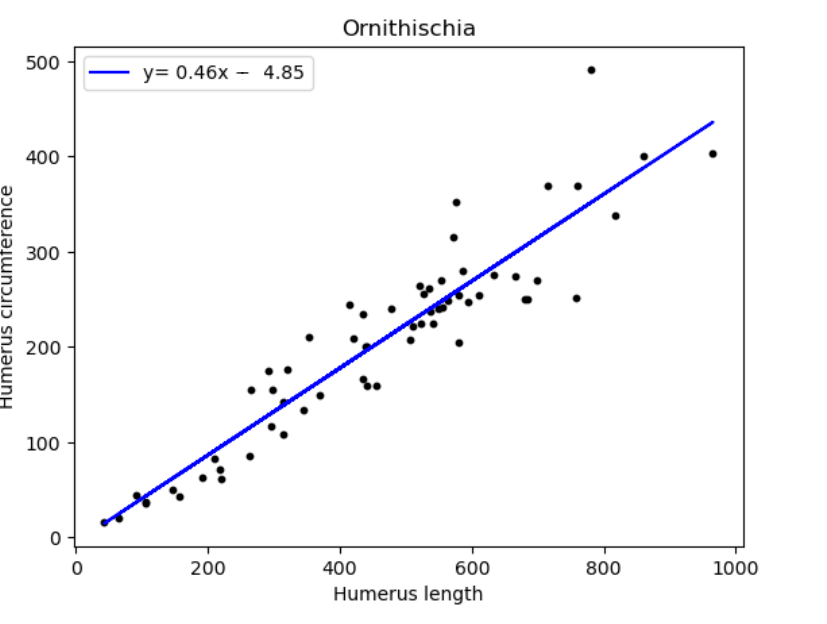}}
\end{minipage}%
\begin{minipage}{.33\linewidth}
\centering
\subfloat[]{\label{main:b}\includegraphics[scale=.3]{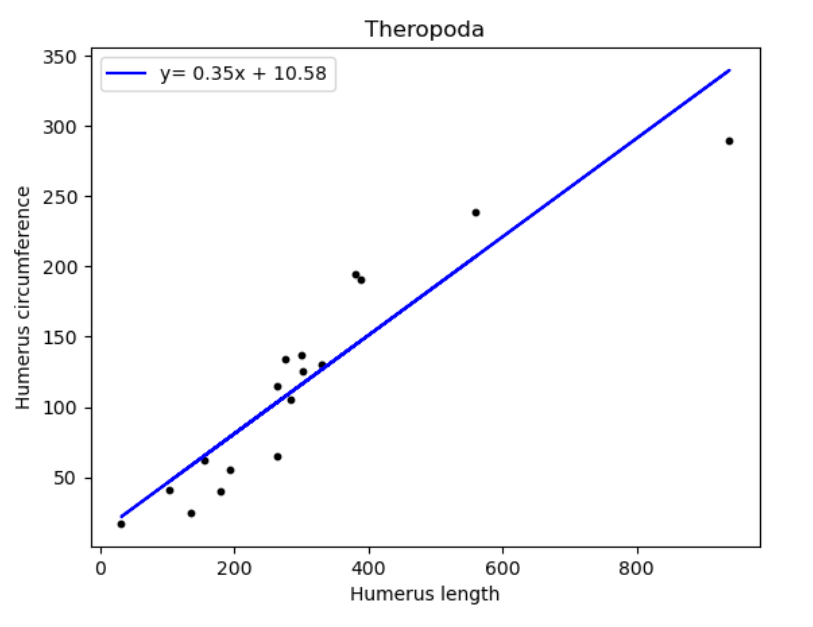}}
\end{minipage}
\begin{minipage}{.33\linewidth}
\centering
\subfloat[]{\label{main:c}\includegraphics[scale=.3]{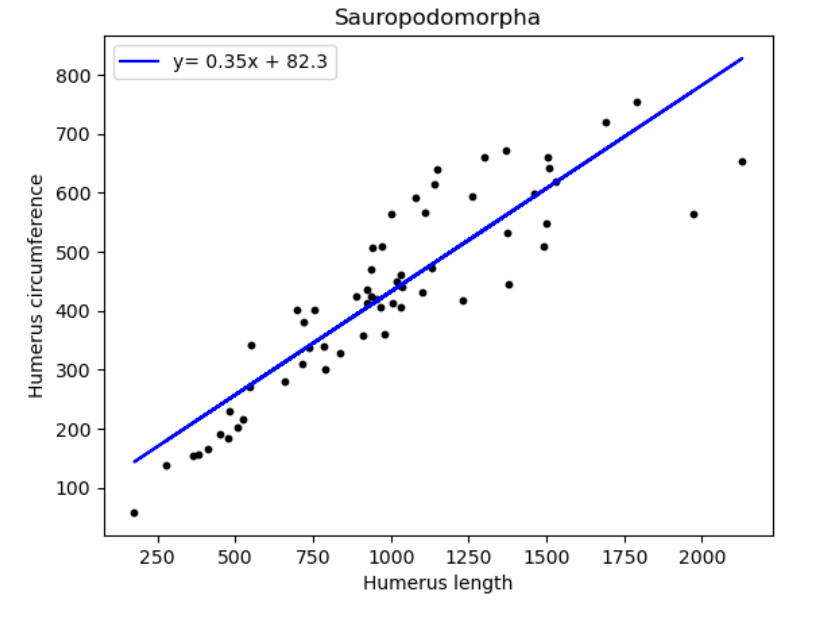}}
\end{minipage}
\caption{(a) Humerus circumference vs. length for Ornithischia (b) Humerus circumference vs. length for Theropoda (c) Humerus circumference vs. length for Sauropodomorpha}
\label{fig:main}
\end{figure}
  \begin{table}[ht]
  \centering
\begin{tabular}{|c|c|c|}
        \hline
        Name of species/group & Slope of line (regression coefficient) & Regression intercept \\
       \hline
         Sauropodomorpha & 0.35 & 82.3 \\
        
       Theropoda & 0.35 & 10.58 \\
      
        Ornithischia & 0.46 & -4.85 \\
        \hline
    \end{tabular}
    \caption{Slope and coefficient values for humerus circumference vs. length}
    \label{tab:simple}
\end{table}
    Looking to confirm any relations between humerus circumference and length amongst the 3 groups. We observe that the same regression coefficient is recorded for Sauropodomorphs and Theropods. The regression coefficient for Ornithischians seem to be substantially greater than both the other species. 
    
From here we can seemingly confirm that both Sauropodomorphs and Theropods are indeed descendants from the same order, by acknowledging that the rate of change of humerus circumference with respect to length was 0.35 for both.

\end{justify}

\begin{flushleft}
    \subsubsection{Tibia circumference with respect to length}
\end{flushleft}
\begin{figure}

\begin{minipage}{.33\linewidth}
\centering
\subfloat[]{\label{main:a}\includegraphics[scale=.3]{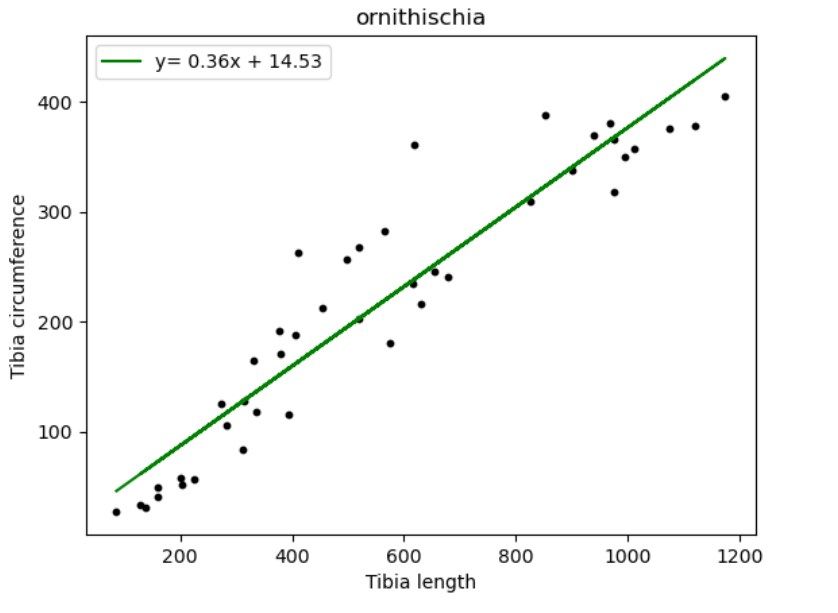}}
\end{minipage}%
\begin{minipage}{.33\linewidth}
\centering
\subfloat[]{\label{main:b}\includegraphics[scale=.3]{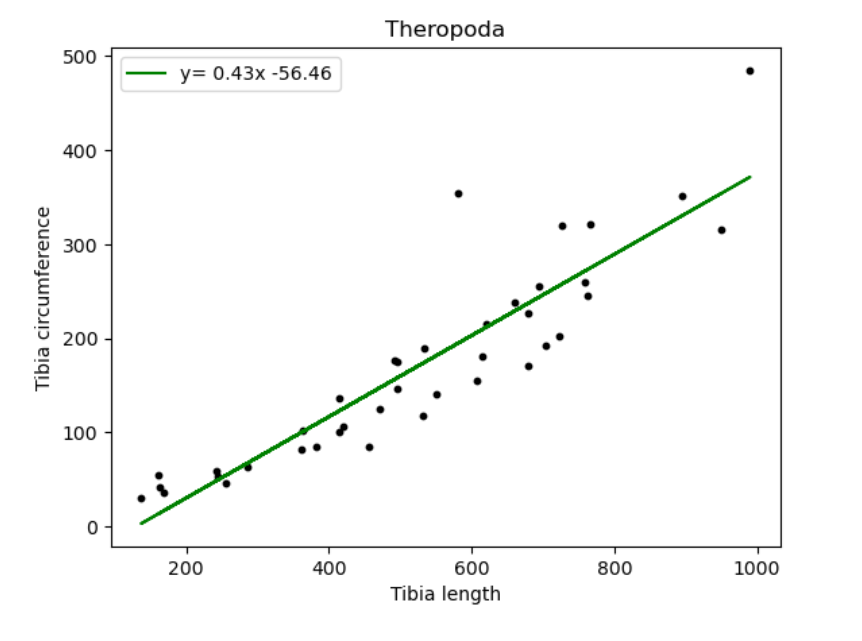}}
\end{minipage}
\begin{minipage}{.33\linewidth}
\centering
\subfloat[]{\label{main:c}\includegraphics[scale=.3]{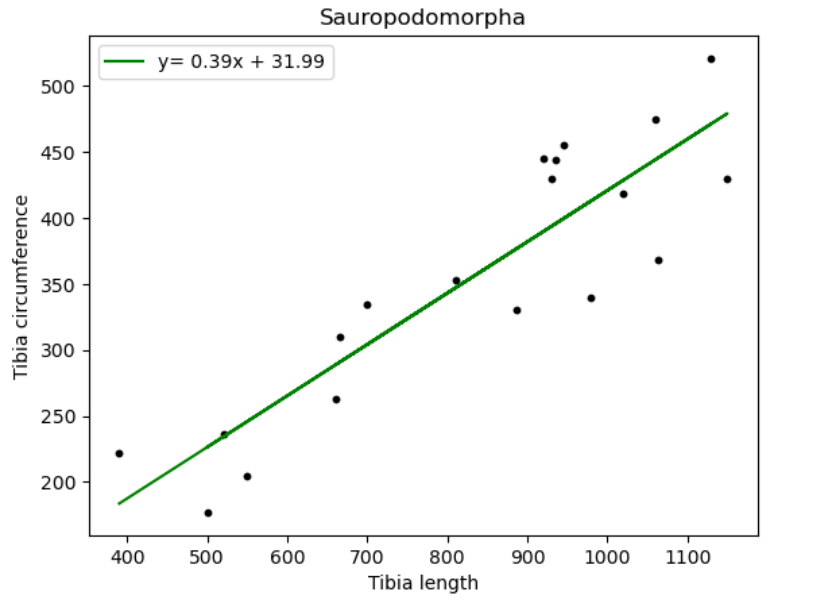}}
\end{minipage}
\caption{(a) Tibia circumference vs. length for Ornithischia (b) Tibia circumference vs. length for Theropoda (c) Tibia circumference vs. length for Sauropodomorpha}
\label{fig:main}
\end{figure}

    \begin{table}[ht]
    \centering
    \begin{tabular}{|c|c|c|}
        \hline
        Name of species/group & Slope of line (regression coefficient) & Regression intercept \\
        \hline
         Sauropodomorpha & 0.39 & 31.99 \\
       
       Theropoda & 0.43 & -56.46 \\
       
        Ornithischia & 0.36 & 14.53 \\
        \hline
    \end{tabular}
    \caption{Slope and coefficient values for tibia circumference vs. length}
    \label{tab:simple}
\end{table}
\begin{justify}
    Finally, taking a look at tibia circumference with respect to length values, we are able to summarise that Theropoda shows the greatest value for increase, that is, 0.43. They are followed by Sauropodomorpha with a 0.39 increase in circumference with length and the smallest increment being shown in Ornithischians.  
    The tibia bone is present in the lower portion of the hind legs of dinosaurs. It is present below the femur. If compared with femur circumference-length ratios. We notice that Theropods show a greater tibia growth rate than femur. This is contradictory to the idea that Theropods are believed to be smaller in size compared to other dinosaur groups due to them being a predator species.  Species adapted to hunting are regarded as smaller and lighter in size (depending on their prey) as opposed to Sauropodomorphs (herbivores) who were known for their considerable size and bulkiness. \cite{botha2022rapid}
\end{justify}
\begin{flushleft}
    \subsubsection{Relation between humerus and femur length}
\end{flushleft}
\begin{figure}
\begin{minipage}{.33\linewidth}
\centering
\subfloat[]{\label{main:a}\includegraphics[scale=.3]{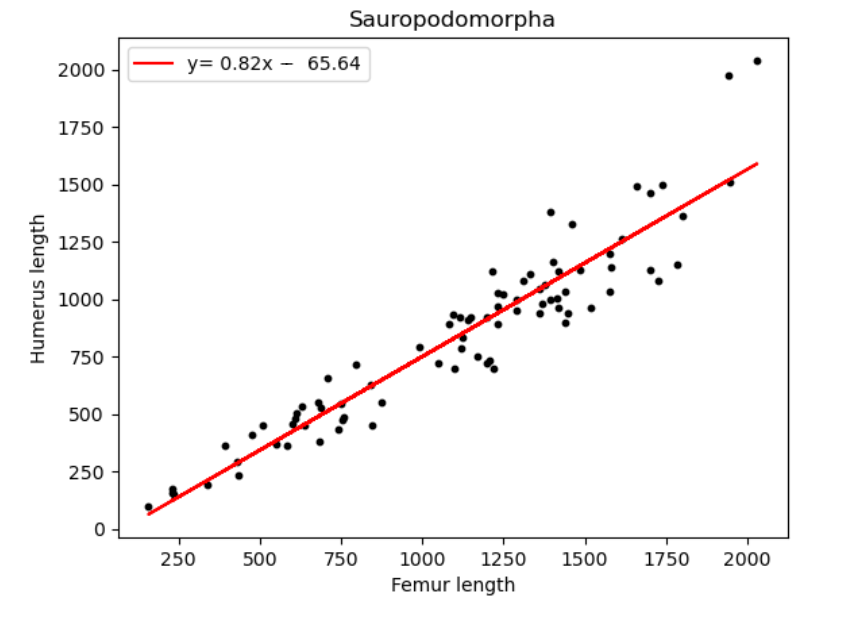}}
\end{minipage}%
\begin{minipage}{.33\linewidth}
\centering
\subfloat[]{\label{main:b}\includegraphics[scale=.3]{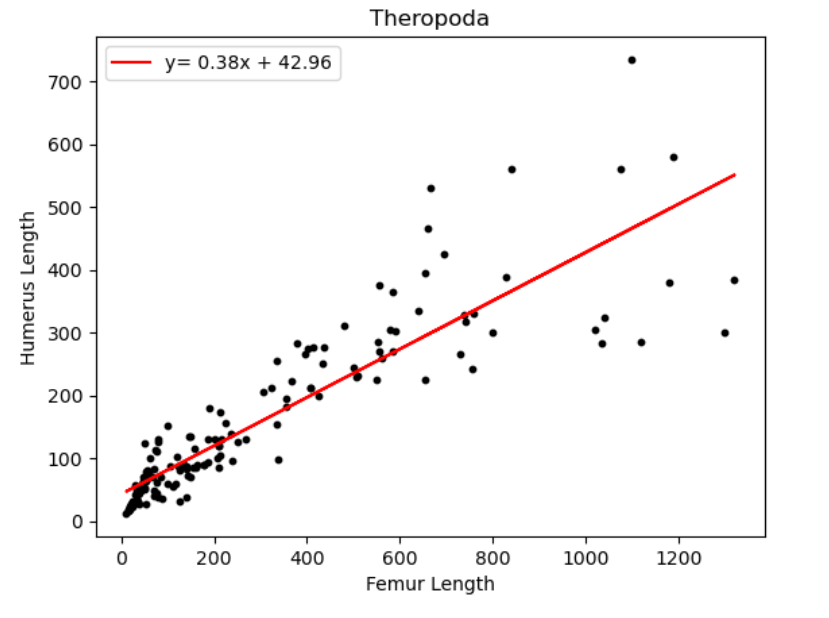}}
\end{minipage}
\caption{(a) Humerus vs. Femur length for Sauropodomorpha (b) Humerus vs. Femur length for Theropoda}
\label{fig:main2}
\end{figure}

\begin{justify}

        \begin{table}[ht]
    \centering
    \begin{tabular}{|c|c|c|}
        \hline
        Name of species/group & Slope of line (regression coefficient) & Regression intercept \\
        \hline
         Sauropodomorpha & 0.82 & -65.64 \\
      
       Theropoda & 0.38 & 42.96 \\
        \hline
    \end{tabular}
    \caption{Slope and coefficient values for humerus vs. femur length}
    \label{tab:simple2}
\end{table}
From Figure \ref{fig:main2} we observe that femur and humerus length ratios derived for Sauropodmorpha and Theropoda show a linear relationship. This can be intuitively explained by simply stating that as humerus length increases, the femur length also proportionally increases to maintain a consistent or steady body size growth\cite{cullen2021growth} and ensure body symmetry. Furthermore, the large value of the slope of humerus-femur length for Sauropodomorphs can be explained by a growth-curve study on Tyrannosauria, particularly the massive Tyrannosaurus (Theropod) \cite{erickson2004gigantism}. The study revealed that a significant increase in growth rates, up to four times, played a pivotal role in the evolution of the impressive size seen in this species. The same study was then conducted for Sauropodomprha which uncovered a similar pattern.\cite{erickson2005assessing}
\end{justify}
\begin{flushleft}
    \subsection{Comparison between all 3 distinct dinosaur species, birds and human}
\end{flushleft}
\begin{justify}
    With this paper, we seek to uncover an unknown relation between 2 biologically unrelated species: Humans and Dinosaurs. Further, we also look to strengthen claims about birds being today's living dinosaur descendants. We attempt to establish these relations by analyzing length-diameter curves. 
\end{justify}
\begin{flushleft}
    \subsubsection{Femur length vs. diameter}
\end{flushleft}
\begin{figure}

\begin{minipage}{.33\linewidth}
\centering
\subfloat[]{\label{main:a}\includegraphics[scale=.3]{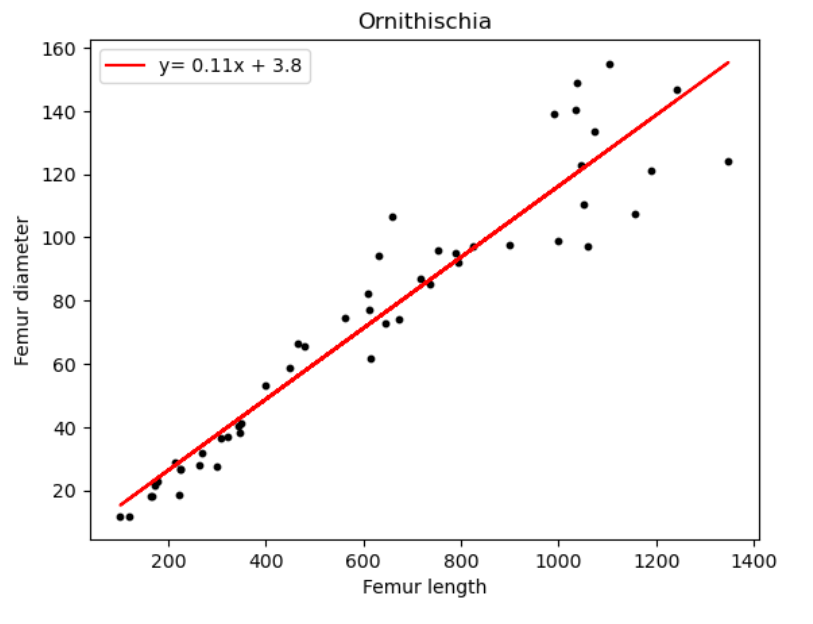}}
\end{minipage}%
\begin{minipage}{.33\linewidth}
\centering
\subfloat[]{\label{main:b}\includegraphics[scale=.3]{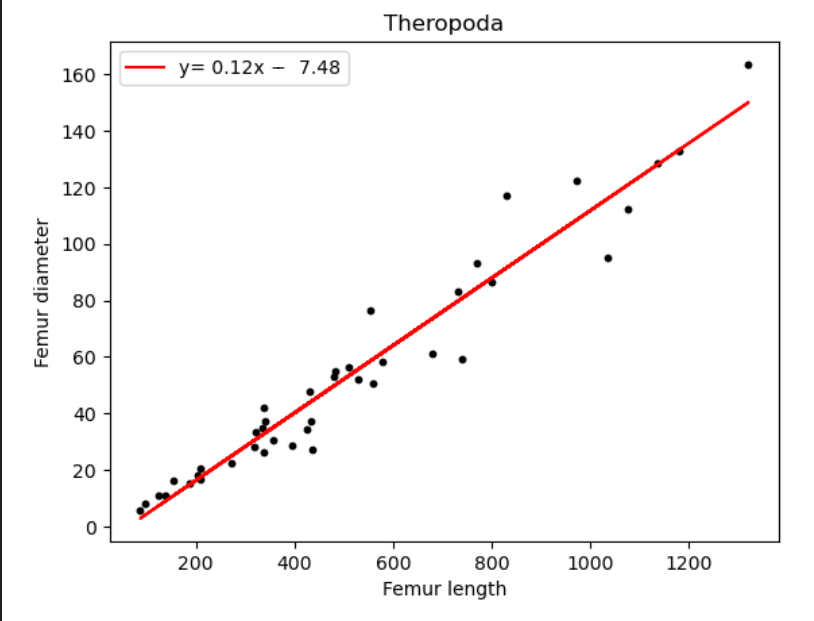}}
\end{minipage}
\begin{minipage}{.33\linewidth}
\centering
\subfloat[]{\label{main:b}\includegraphics[scale=.3]{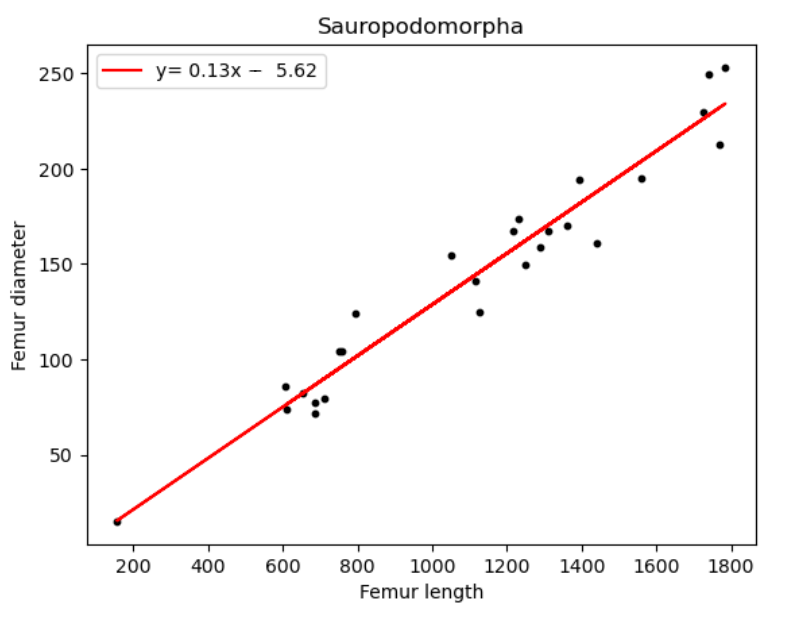}}
\end{minipage}
\begin{minipage}{.33\linewidth}
\centering
\subfloat[]{\label{main:b}\includegraphics[scale=.3]{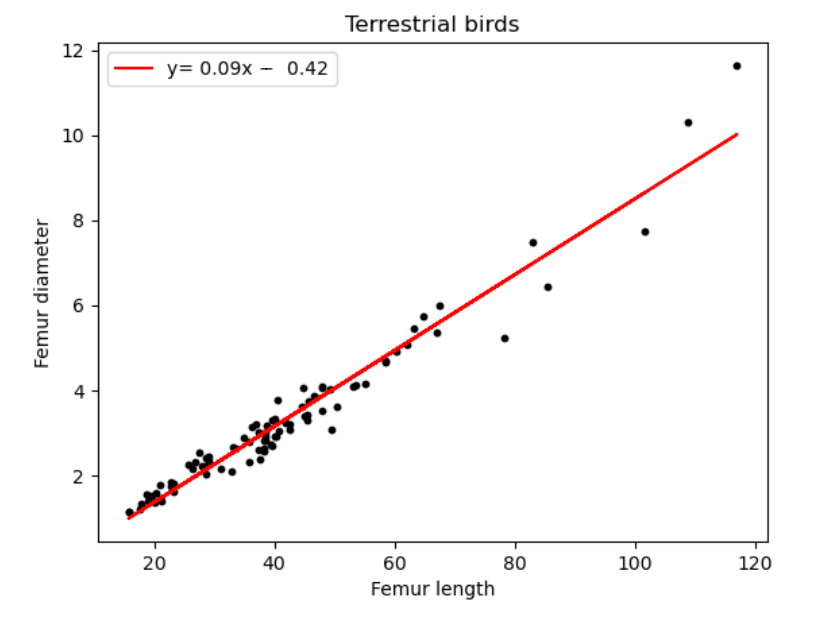}}
\end{minipage}
\begin{minipage}{.33\linewidth}
\centering
\subfloat[]{\label{main:c}\includegraphics[scale=.3]{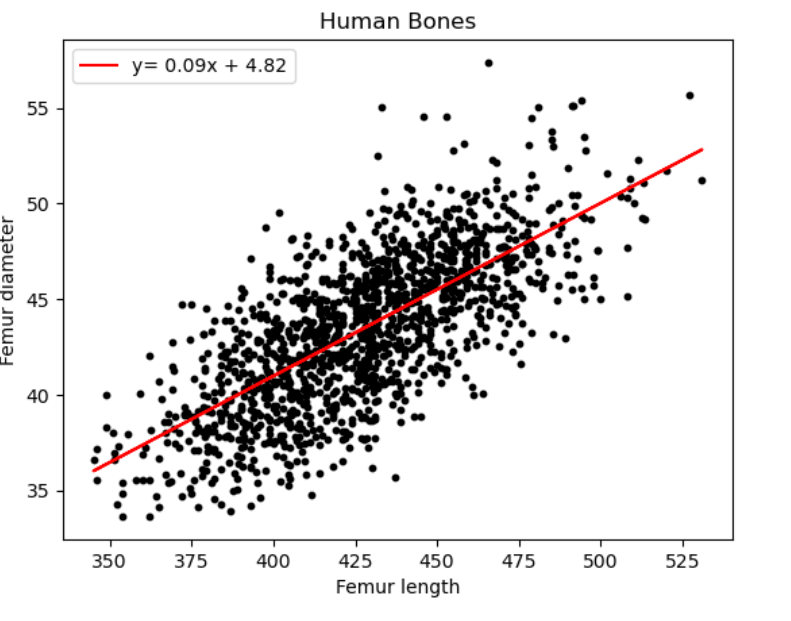}}
\end{minipage}
\caption{(a) Femur diameter vs. length for Ornithischia (b) Femur diameter vs. length for Theropoda (c) Femur diameter vs. length for Sauropodomorpha (d) Femur diameter vs. length for terrestrial birds (incl. wading birds) (e) Femur diameter vs. length for humans}
\label{diavslen}
\end{figure}
 \begin{table}[ht]
    \centering
    \begin{tabular}{|c|c|c|}
        \hline
        Name of species/group & Slope of line (regression coefficient) & Regression intercept \\
        \hline
        Sauropodomorpha & 0.13 & -5.62 \\
        Theropoda & 0.12 & -7.48 \\
        Ornithischia & 0.11 & 3.8 \\
        Humans & 0.09 & 4.82 \\
        Terrestrial Birds & 0.09 & -0.42 \\
        \hline
    \end{tabular}
    \caption{Slope and coefficient values for femur length vs. diameter}
    \label{tab:diavslen}
\end{table}
\begin{justify}
    
    We can deduce from Figure \ref{diavslen} and Table \ref{tab:diavslen}, a minimal value distinction for all above species. All the dinosaur taxa ratios differ from each other only by 0.1, with Sauropodomorpha being the highest at 0.13. The most unexpected result however is humans and birds having equal coefficient values.
    
    For long, the origin of the bird puzzled scientists. However, today, the Theropod hypothesis is almost universally accepted which links them to be descendants to carnivorous dinosaurs such as Deinonychus, Velociraptor, Troodon and Oviraptors which share a great deal of similarities with birds\cite{chiappe2009downsized}. However, this is inconsistent with our values, which do not indicate a relationship between Theropods and birds. 
   
    We also notice a close ratio between Ornithischians and birds, this may be due to their shared ancestry and several osteological similarities such as homologous pelvic bones (a retroverted pubis and elongated iliac). A study also concluded convergence between birds and basal ornithischians, which further affirms this ratio.\cite{bates2012computational}
\end{justify}
\begin{flushleft}
    \subsubsection{Humerus length vs. diameter}
\end{flushleft}
\begin{justify}
\begin{figure}
\begin{minipage}{.33\linewidth}
\centering
\subfloat[]{\label{main:a}\includegraphics[scale=.3]{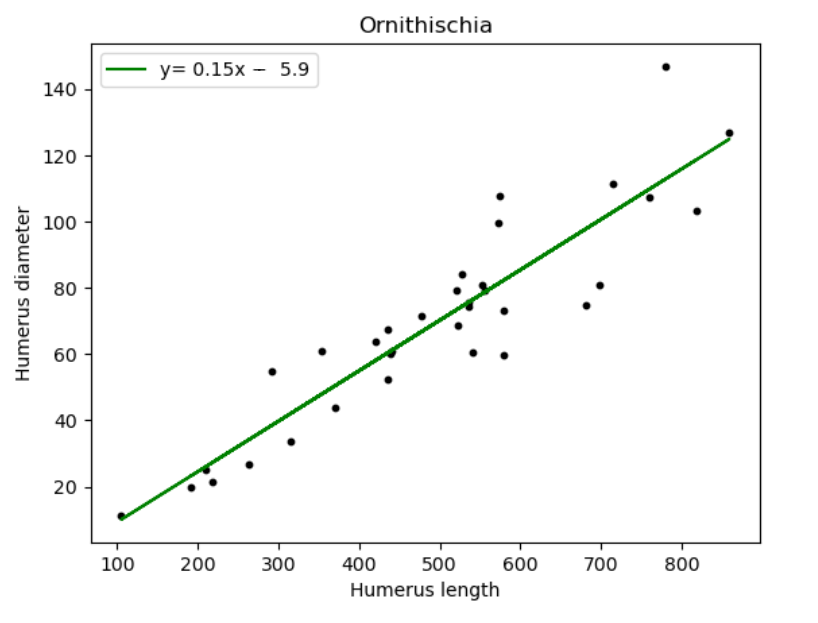}}
\end{minipage}%
\begin{minipage}{.33\linewidth}
\centering
\subfloat[]{\label{main:b}\includegraphics[scale=.3]{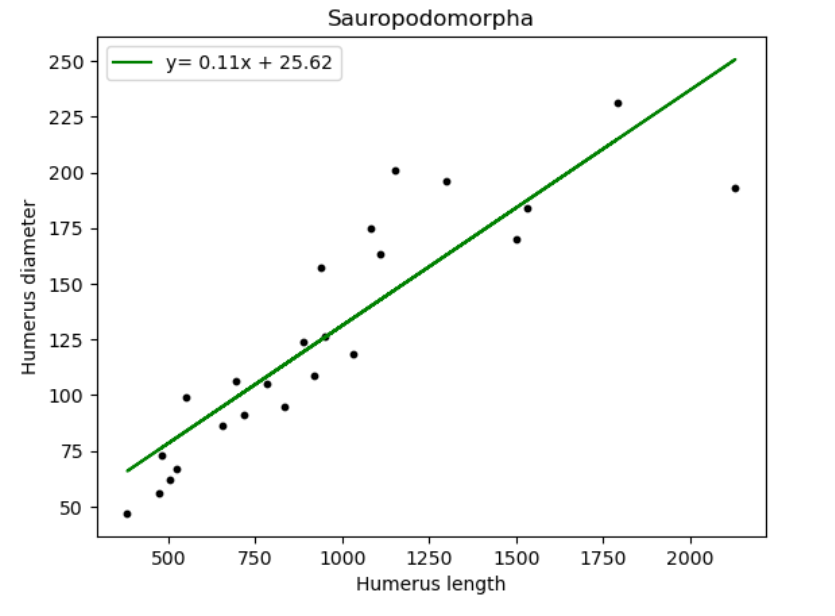}}
\end{minipage}
\begin{minipage}{.33\linewidth}
\centering
\subfloat[]{\label{main:c}\includegraphics[scale=.3]{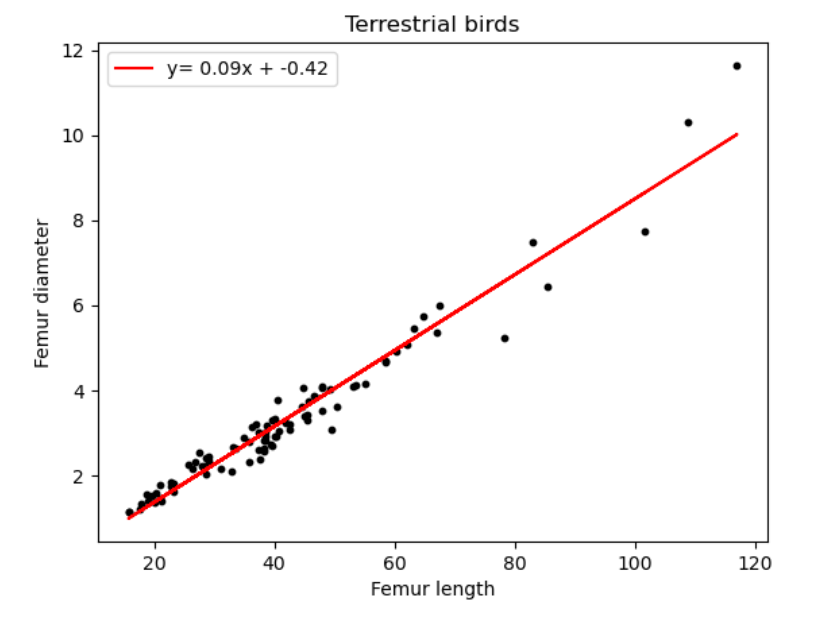}}
\end{minipage}
\begin{minipage}{.33\linewidth}
\centering
\subfloat[]{\label{main:c}\includegraphics[scale=.3]{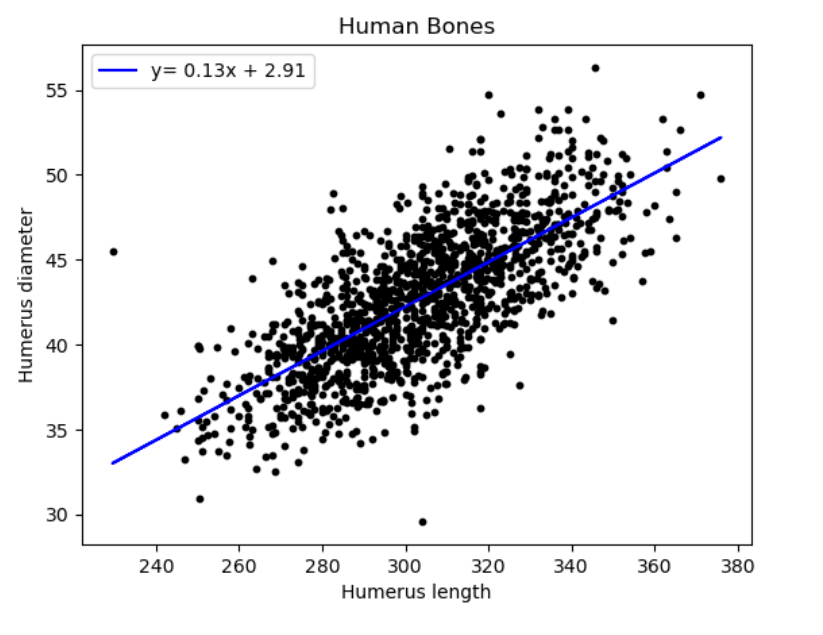}}
\end{minipage}
\caption{(a) Humerus diameter vs. length for Ornithischia (b) Humerus diameter vs. length for Sauropodomorpha (c) Humerus diameter vs. length for terrestrial birds (incl. wading birds) (d) Humerus diameter vs. length for humans}
\label{fig:main}
\end{figure}
\begin{table}[ht]
    \centering
    \begin{tabular}{|c|c|c|}
        \hline
        Name of species/group & Slope of line (regression coefficient) & Regression intercept \\
        \hline
        Sauropodomorpha & 0.11 & 25.62 \\
        Ornithischia & 0.15 & -5.9 \\
        Humans & 0.13 & 2.91 \\
        Terrestrial Birds & 0.05 & 1.08 \\
        \hline
    \end{tabular}
    \caption{Slope and coefficient values for humerus length vs. diameter}
    \label{tab:humerlenvsdia}
\end{table}
Comparing the increase in humerus length with respect to diameter in Table \ref{tab:humerlenvsdia}, our results vary. 

We see ornithischians having the highest slope of 0.15, which is consistent with the high value of humerus circumference-length ratio, suggesting that overall Ornithischians saw the largest rate of increase in forelimb bones (humerus, being the largest of them). 

Birds seem to have the lowest ratio of 0.05, which we believe can be attributed to their small size as well as their diets (nutrients and amount consumed). 

As for Sauropods, the humerus length in relation to diameter (0.11) is still less than the femur diameter in relation to length (0.13). This follows a similar trend in circumference- length graphs as well, further alluding to the connection between 3 linear quantities of length, circumference and diameter. 
\end{justify}
\begin{flushleft}
    \subsection{Deeper dive into biological connections between human and bird}
\end{flushleft}
\begin{justify}
    To answer for the similarities in values obtained for human and bird bone comparisons, we turn to the theory of convergent evolution\cite{seed2009intelligence}: the phenomenon of different species developing similar features or traits while evolving simultaneously. Here, the traits acquired by the species are not a result of shared ancestry but are rather due to external factors, such as environment or survival pressure. 

    A study conducted on intelligence in corvids (large-brained birds) and apes details how enlarged brain sizes in both corvids and apes could have been a result of convergent evolution and verifies the resembling environmental conditions faced by both species. From this study, the dietary overlap between the 2 species is of relevance to us. Both corvids and primates displayed an evolution of intelligence when they engaged in extractive foraging, specifically as omnivores. Similar diets suggest the availability of the same nutrients and proteins which we can articulate as proportional bone growth in both species. However, no detailed experiments or studies have been carried out on relationship between convergent evolution and body size growth.\cite{seed2009intelligence} \cite{seeley1901dragons} 
\section{Conclusion}
    We performed linear regression analysis on bone data of Ornithischia, Theropoda and Sauropodomorpha. We found the slopes, which are a measure of the ratios of the average linear trend for length vs circumference or length vs diameter for various bone data available. We see similar trends in these ratios, hinted by very comparable slopes of these graphs for only dinosaur species. This establishes that these ratios must be similar for similar species of Animalia. This can be understood by similarities in evolution, environments and genetic makeup, leading to similar bone structures.

    We further compare similar ratios for terrestrial birds and humans in contrast with the dinosaur data. We see comparable in the trends of human and bird data.

    We believe that at some point in time, humans and terrestrial birds did share similar environmental conditions which may have led to a convergent evolution at least in terms of their bone ratios. The data for dinosaurs suggest a strong relation in the femur bone ratios, which is not seen between dinosaurs and birds as expected by widely popular theories. The key result which we wish to convey is that these ratios between humans and terrestrial birds show much similar trend in femur bone structure. The humerus being the forelimb bone, may have evolved at a different time due to change in environmental conditions forcing the terrestrial birds to grow wings. This explains the stark difference between the humerus ratios of birds and humans. Even despite of this, the humerus ratios are nowhere near to that of dinosaurs, hinting at more rich data science questions in this field.

\section{Acknowledgements}

We would like to acknowledge the authors of \cite{plos} for providing the data which is used in this work.

\section{Appendix I}
The table \ref{goodness} shows the goodness of the fit for the various data sets available. Figures \ref{fig:femlen} to \ref{fig:tibcircum} show the distribution of data as histograms.

\begin{table}[h!]
\begin{tabular}{|l|l|l|}
\hline
                & Relation                           & R-squared \\ \hline
Ornithischia    & Femur length vs. circumference     & 0.937     \\ 
                & Tibia length vs. circumference     & 0.885     \\ 
                & Humerus length vs.   circumference & 0.866     \\ e
                & Femur length vs. diameter          & 0.907     \\ 
                & Humerus length vs. diameter        & 0.828     \\ \hline
Theropoda       & Femur length vs. circumference     & 0.959     \\ 
                & Tibia length vs. circumference     & 0.815     \\ 
                & Humerus length vs.   circumference & 0.849     \\ 
                & Femur vs. Humerus length           & 0.790     \\ 
                & Femur length vs. diameter          & 0.942     \\ \hline
Sauropodomorpha & Femur length vs. circumference     & 0.901     \\ 
                & Tibia length vs. circumference     & 0.785     \\ 
                & Humerus length vs.   circumference & 0.792     \\ 
                & Humerus vs. Femur    length        & 0.885     \\ 
                & Femur length vs. diameter          & 0.946     \\ 
                & Humerus length vs. diameter        & 0.784     \\ \hline
Birds           & Humerus length vs. diameter        & 0.743     \\ 
                & Femur length vs. diameter          & 0.952     \\ 
                & Tibia length vs. diameter          & 0.837     \\ \hline
Humans          & Humerus length vs. diameter        & 0.534     \\ 
                & Femur length vs. diameter          & 0.513     \\ 
                & Tibia length vs. diameter          & 0.276     \\ \hline
\end{tabular}
    \caption{R-squared values for various fits. The closer the R-squared is to 1, the better is the fit.}
    \label{goodness}
\end{table}

\begin{figure}[h!]
\begin{minipage}{.33\linewidth}
\centering
\subfloat[]{\label{main:a}\includegraphics[scale=.3]{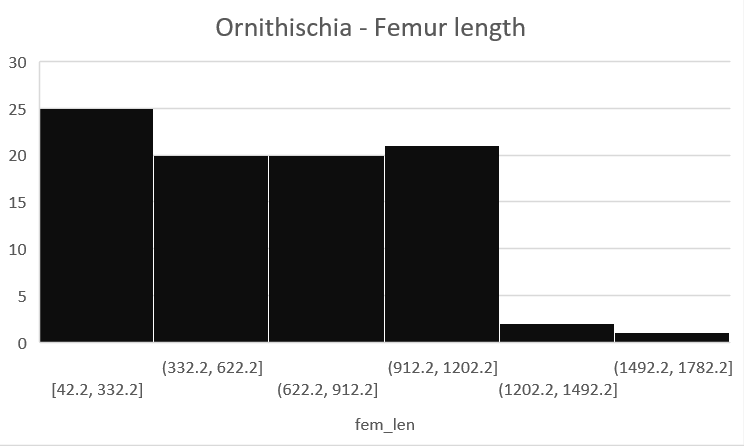}}
\end{minipage}%
\begin{minipage}{.33\linewidth}
\centering
\subfloat[]{\label{main:b}\includegraphics[scale=.3]{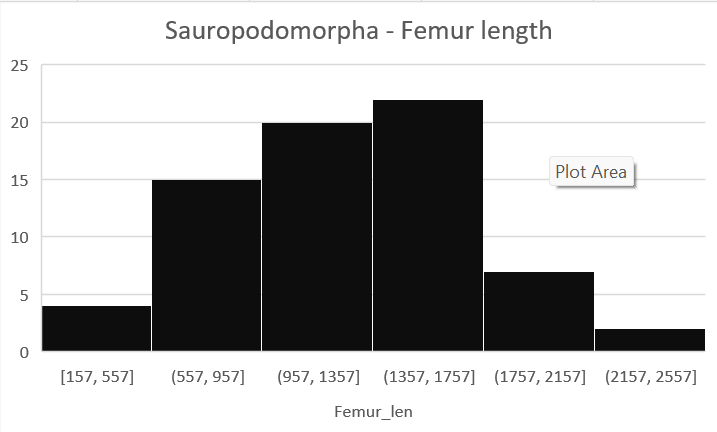}}
\end{minipage}
\begin{minipage}{.33\linewidth}
\centering
\subfloat[]{\label{main:c}\includegraphics[scale=.3]{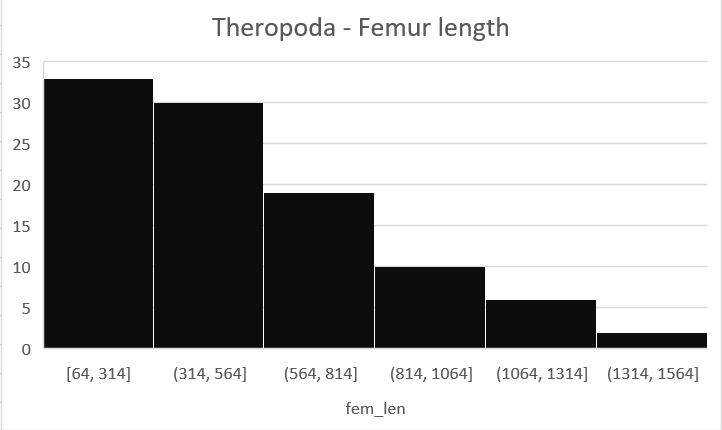}}
\end{minipage}
\caption{Femur lengths}
\label{fig:femlen}
\end{figure}

\begin{figure}[h!]
\begin{minipage}{.33\linewidth}
\centering
\subfloat[]{\label{main:a}\includegraphics[scale=.3]{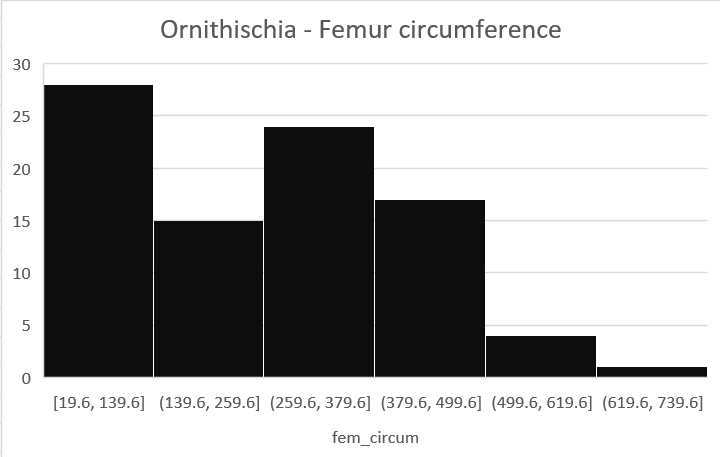}}
\end{minipage}%
\begin{minipage}{.33\linewidth}
\centering
\subfloat[]{\label{main:b}\includegraphics[scale=.3]{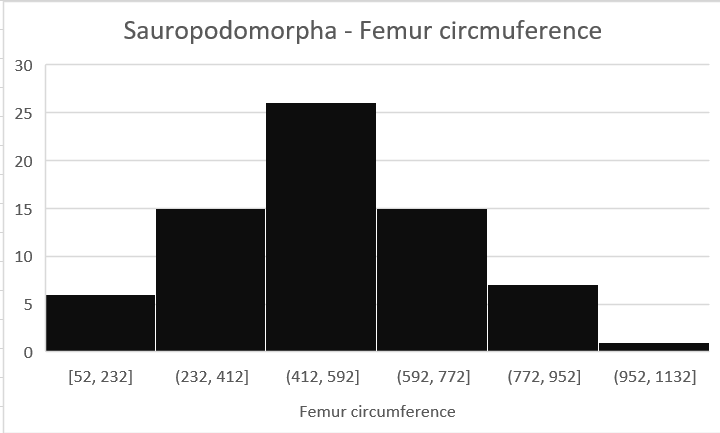}}
\end{minipage}
\begin{minipage}{.33\linewidth}
\centering
\subfloat[]{\label{main:c}\includegraphics[scale=.3]{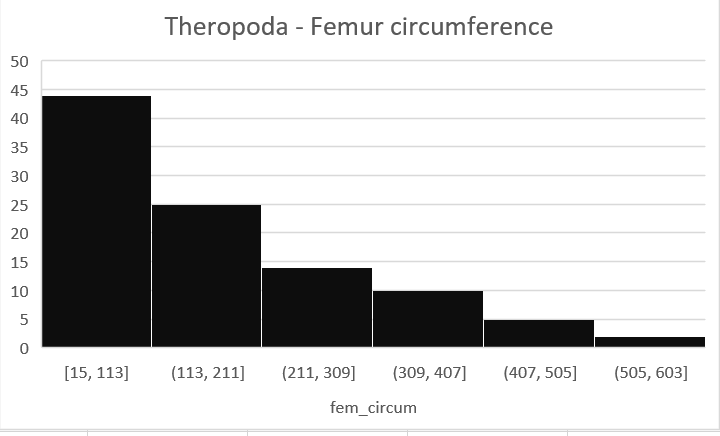}}
\end{minipage}
\caption{Femur circumferences}
\label{fig:femcircum}
\end{figure}

\begin{figure}[h!]
\begin{minipage}{.33\linewidth}
\centering
\subfloat[]{\label{main:a}\includegraphics[scale=.3]{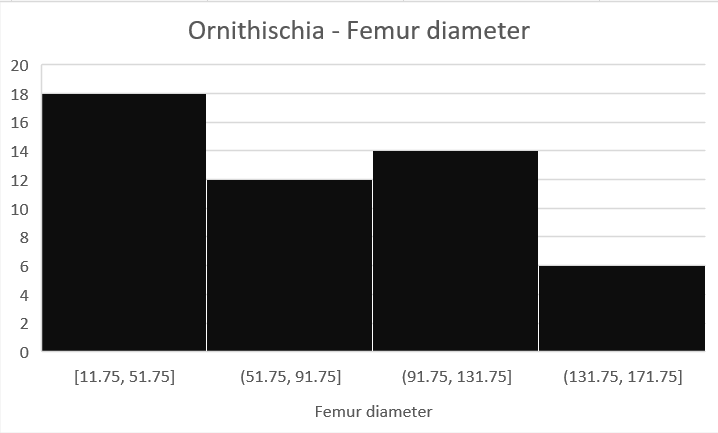}}
\end{minipage}%
\begin{minipage}{.33\linewidth}
\centering
\subfloat[]{\label{main:b}\includegraphics[scale=.3]{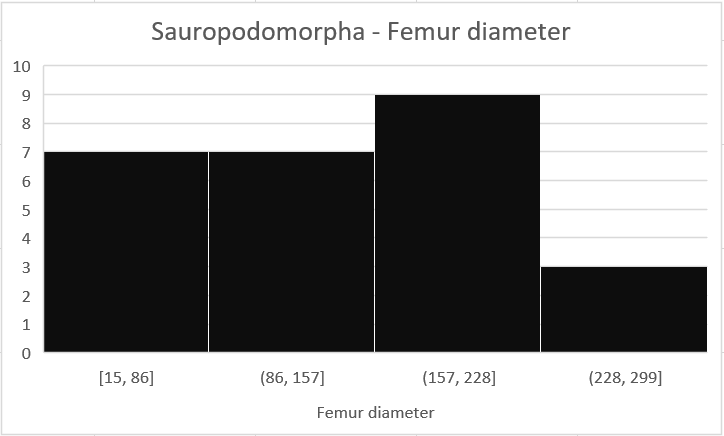}}
\end{minipage}
\begin{minipage}{.33\linewidth}
\centering
\subfloat[]{\label{main:c}\includegraphics[scale=.3]{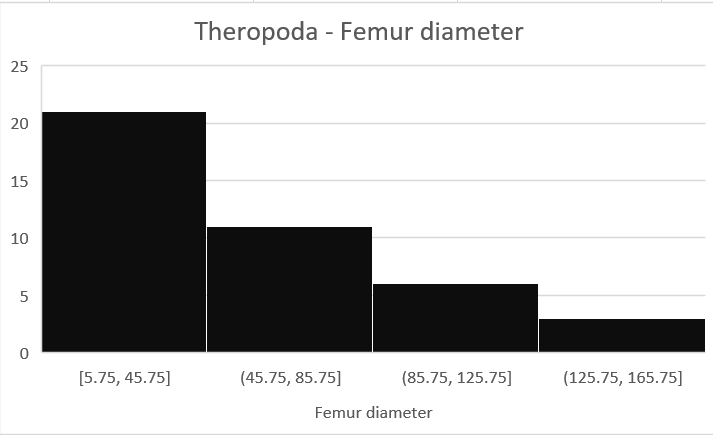}}
\end{minipage}
\caption{Femur diameters}
\label{fig:femdia}
\end{figure}

\begin{figure}[h!]
\begin{minipage}{.33\linewidth}
\centering
\subfloat[]{\label{main:a}\includegraphics[scale=.3]{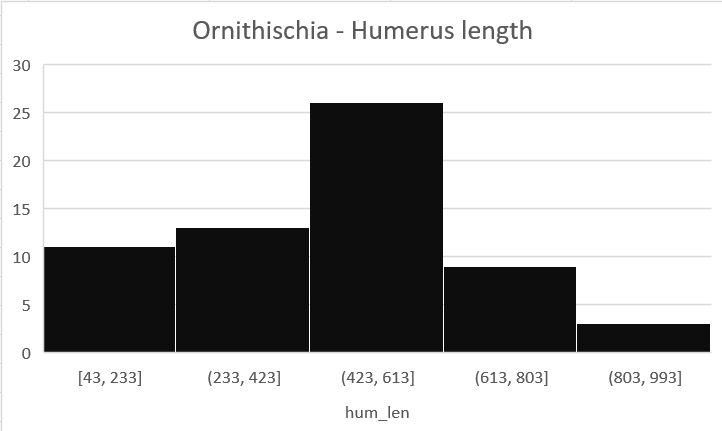}}
\end{minipage}%
\begin{minipage}{.33\linewidth}
\centering
\subfloat[]{\label{main:b}\includegraphics[scale=.3]{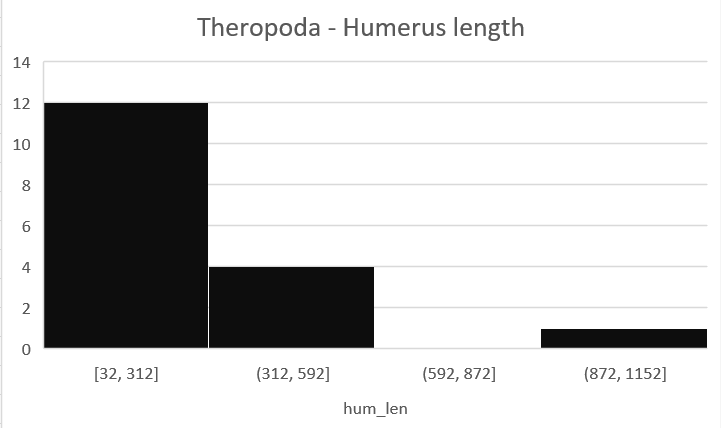}}
\end{minipage}
\caption{Humerus lengths}
\label{fig:humlen}
\end{figure}

\begin{figure}[h!]
\begin{minipage}{.33\linewidth}
\centering
\subfloat[]{\label{main:a}\includegraphics[scale=.3]{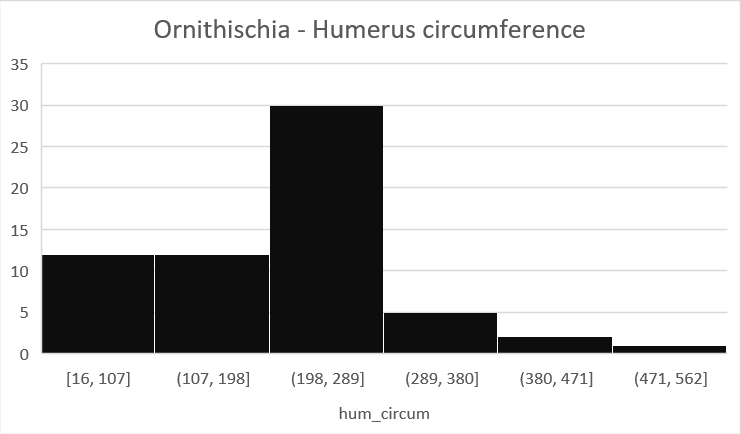}}
\end{minipage}%
\begin{minipage}{.33\linewidth}
\centering
\subfloat[]{\label{main:b}\includegraphics[scale=.3]{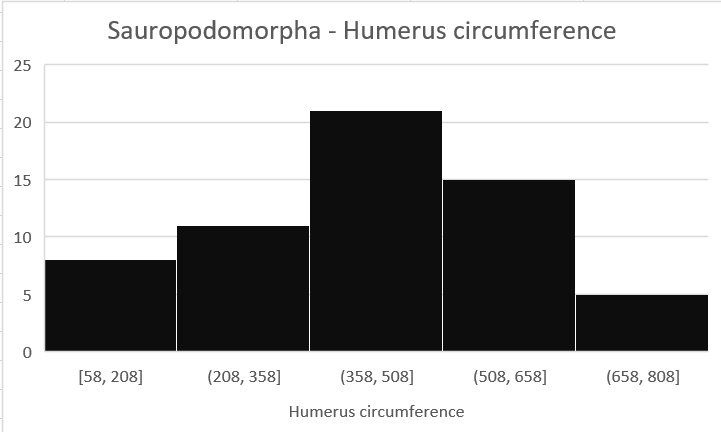}}
\end{minipage}
\begin{minipage}{.33\linewidth}
\centering
\subfloat[]{\label{main:c}\includegraphics[scale=.3]{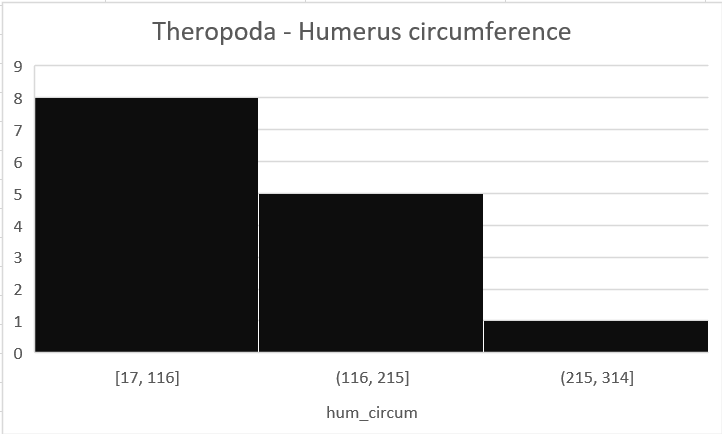}}
\end{minipage}
\caption{Humerus circumference}
\label{fig:humcircum}
\end{figure}

\begin{figure}[h!]
\begin{minipage}{.33\linewidth}
\centering
\subfloat[]{\label{main:a}\includegraphics[scale=.3]{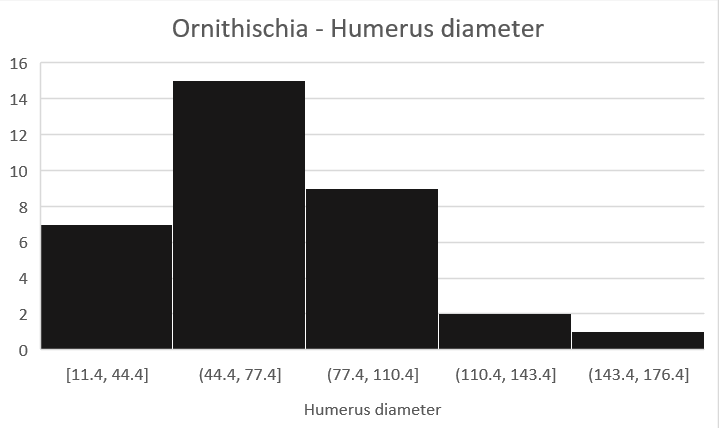}}
\end{minipage}%
\begin{minipage}{.33\linewidth}
\centering
\subfloat[]{\label{main:b}\includegraphics[scale=.3]{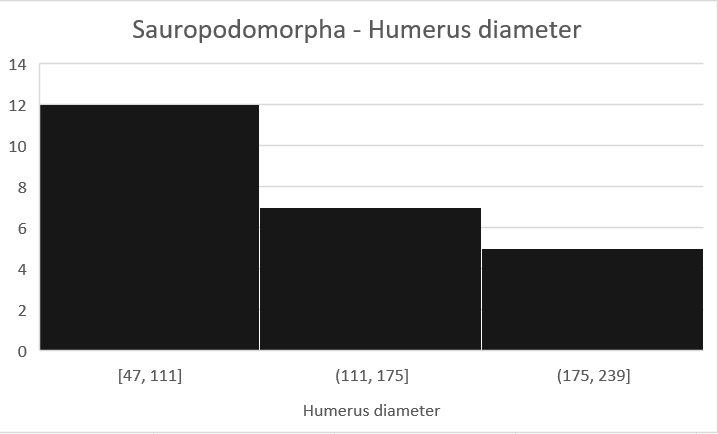}}
\end{minipage}
\caption{Humerus diameter}
\label{fig:humdia}
\end{figure}

\begin{figure}[h!]
\begin{minipage}{.33\linewidth}
\centering
\subfloat[]{\label{main:a}\includegraphics[scale=.3]{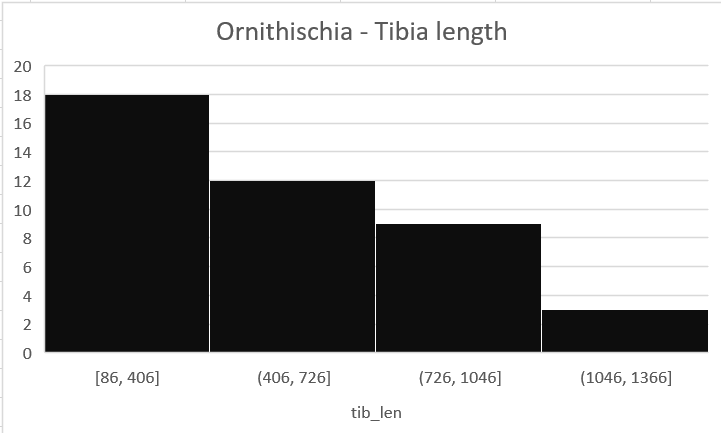}}
\end{minipage}%
\begin{minipage}{.33\linewidth}
\centering
\subfloat[]{\label{main:b}\includegraphics[scale=.3]{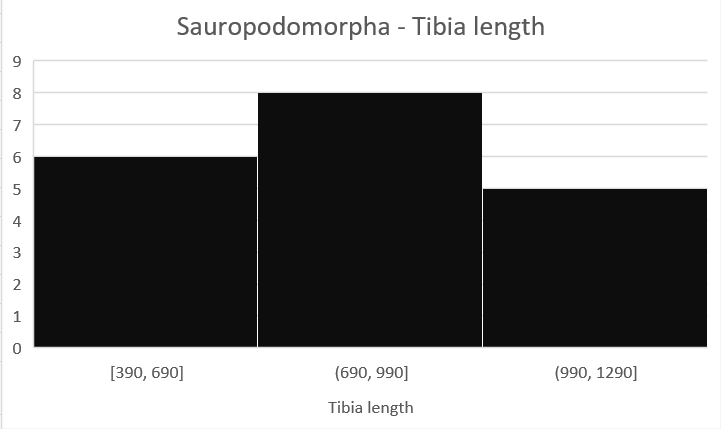}}
\end{minipage}
\begin{minipage}{.33\linewidth}
\centering
\subfloat[]{\label{main:c}\includegraphics[scale=.3]{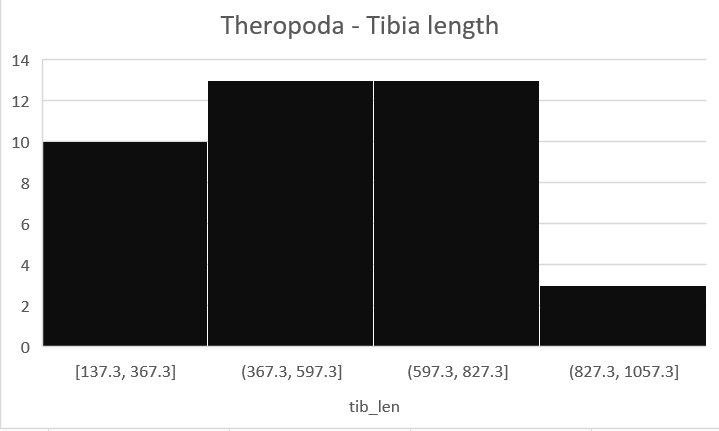}}
\end{minipage}
\caption{Tibia lengths}
\label{fig:tibailen}
\end{figure}

\begin{figure}[h!]
\begin{minipage}{.33\linewidth}
\centering
\subfloat[]{\label{main:a}\includegraphics[scale=.3]{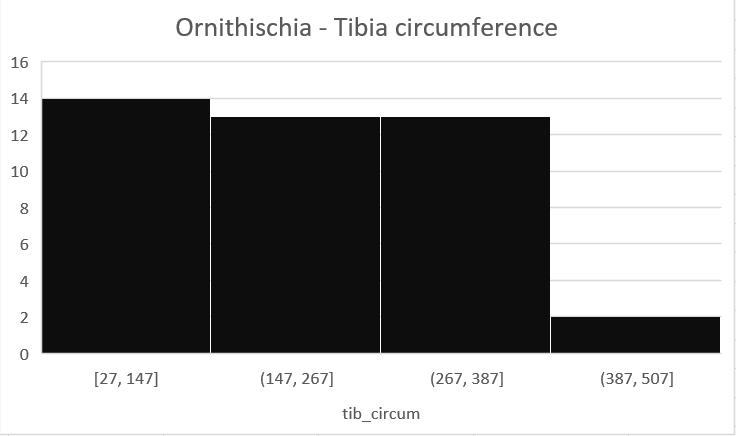}}
\end{minipage}%
\begin{minipage}{.33\linewidth}
\centering
\subfloat[]{\label{main:b}\includegraphics[scale=.3]{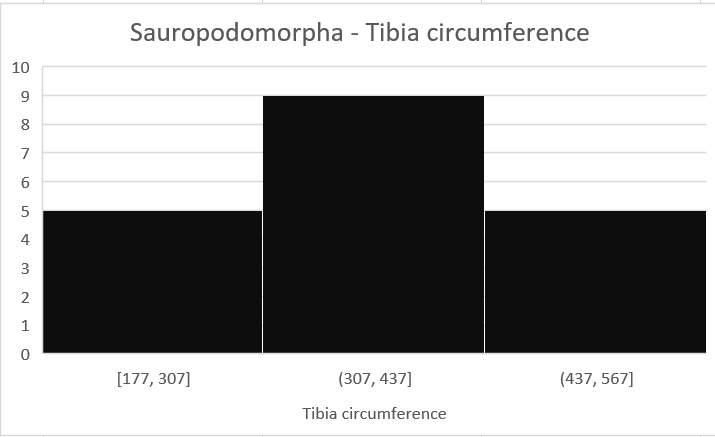}}
\end{minipage}
\begin{minipage}{.33\linewidth}
\centering
\subfloat[]{\label{main:c}\includegraphics[scale=.3]{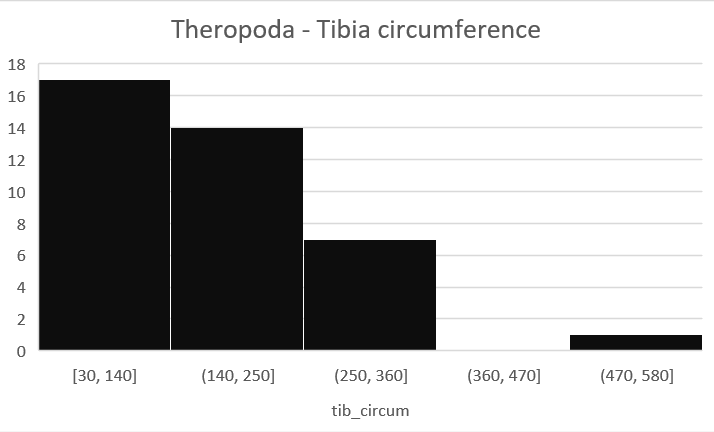}}
\end{minipage}
\caption{Tibia circumferences}
\label{fig:tibcircum}
\end{figure}

\end{justify}
\clearpage
\bibliographystyle{unsrt}
\bibliography{bib.bib}
\end{document}